\documentstyle[epsf,12pt,aasms4]{article}

\begin{document}
\newcommand{\apgt}{\ {\raise-.5ex\hbox{$\buildrel>\over\sim$}}\ }
\newcommand{\aplt}{\ {\raise-.5ex\hbox{$\buildrel<\over\sim$}}\ }

\title{How Rare Are Extraterrestrial Civilizations and When Did They
Emerge?}
\author{Mario Livio\\
Space Telescope Science Institute\\
3700 San Martin Drive\\
Baltimore, MD 21218}

\begin{abstract}

It is shown that, contrary to an existing claim, the near equality
between the 
lifetime of the sun and the timescale of biological evolution on earth
does not 
necessarily imply that extraterrestrial civilizations are exceedingly
rare.  
Furthermore, on the basis of simple assumptions it is demonstrated that
a near 
equality between these two timescales may be the most probable relation.
A calculation of the cosmic history of carbon production which is based
on the 
recently determined history of the star formation rate suggests that the
most 
likely time for intelligent civilizations to emerge in the universe, was
when 
the universe was already older then about 10~billion years (for an
assumed 
current age of about 13~billion years).

\end{abstract}

\section{Introduction}

With the recent discovery of several extra-solar planets (and brown
dwarfs) 
around solar type stars (e.g.\ Mayor \& Queloz 1995; Buter \& Marcy
1996; 
Basri, Marcy \& Graham 1996) the question of the potential existence of 
extraterrestrial ``intelligent'' civilizations has become more
intriguing than 
ever.  While this topic has been the subject of extensive speculations
and many 
ill-defined (often by necessity) probability estimates, at least one
study 
(Carter 1983) has examined it from a more global, statistical
perspective.  
That study concluded, on the basis of the near equality between the
timescale 
of biological evolution on Earth, $\tau_l$, and the lifetime of the Sun, 
$\tau_\odot$, that extraterrestrial civilizations are exceedingly rare,
even 
if conditions favorable for the development of life are relatively
common.  

The conclusion on the rarity of extraterrestrial intelligent
civilizations 
(Carter 1983; and see also Barrow \& Tipler 1986) was based on one
crucial 
{\it assumption\/} and one {\it observation\/}.  The {\it assumption\/}
is 
that the timescale of biological evolution on a given planet, $\tau_l$,
and 
the lifetime of the central star, $\tau_*$, are a~priori entirely
independent 
quantities. Put differently, this assumes that intelligent life forms at
some 
random time with respect to the main sequence lifetime of the star.  The 
{\it observation\/} is that in the earth-sun system $\tau_l \sim \tau_*$
(to 
within a factor 2; for definiteness I will take from now on $\tau_l$ to 
represent the timescale for the appearance of land life). For
completeness, I 
will reproduce here the argument briefly.  {\it If\/} $\tau_l$ and
$\tau_*$ 
are indeed independent quantities, then most probably either $\tau_l \gg 
\tau_*$ or $\tau_l \ll \tau_*$ (the set of $\tau_l \sim \tau_*$ is of
very 
small measure for two independent quantities).  If, however, $\tau_l \ll 
\tau_*$ {\it generally}, then it is very difficult to understand why in
the 
first system found to exhibit an intelligent civilization (the earth-sun 
system), it was found that $\tau_l \sim \tau_*$.  If, on the other hand, 
{\it generally\/} $\tau_l \gg \tau_*$, then it is clear that the first
system 
found to contain an intelligent civilization is likely to have $\tau_l
\sim 
\tau_*$ (since for $\tau_l \gg \tau_*$ a civilization would not have
developed).  
Thus, according to this argument, one has to conclude that typically
$\tau_l 
\gg \tau_*$, namely, that {\it generally\/} intelligent civilizations
will not 
develop, and that the earth is the extremely rare exception.  What I
intend 
first to show in the present work is that this conclusion is at best
premature, 
by demonstrating that not only that $\tau_l$ and $\tau_*$ may not be 
independent, but also that $\tau_l/\tau_* \sim 1$ may in fact be the
{\it most 
probable\/} value for this ratio. This is done in \S~2.  In \S~3 I use
the 
recently determined cosmic star formation history to estimate the most
likely 
time for intelligent civilizations to emerge in the universe.

\section{The Relation Between $\tau_l$ and $\tau_*$}

Superficially it appears that $\tau_l$ (which is determined mainly by 
biochemical reactions and evolution of species) and $\tau_*$ (which is
determined 
by nuclear burning reactions) should be independent.  It suffices to
note, 
however, that light energy (from the central star) exceeds by 2--3
orders of 
magnitude all other sources of energy that can drive chemical evolution
in the 
prebiotic environment (e.g.\ Deamer 1997), to realize that $\tau_l$ may
in fact 
depend on $\tau_*$ (note that the statement that intelligent life will
not 
develop for $\tau_l \gg \tau_*$ also constitutes a qualitative
dependence).  
Below I identify a specific physical process that can, in principle at
least, 
relate the two timescales.  First I would like however to point out the 
following {\it general\/} property.  Imagine that we find that the ratio
$\tau_l/\tau_*$ 
can be described by some function of the form $\tau_l/\tau_* =
f(\tau_*)$, and 
that we further find that the function $f$ is monotonically increasing
(at least 
for the narrow range of values of $\tau_*$ corresponding to stars which
allow 
the development of life, see below).  This situation is shown
schematically in 
Fig.~1.  In such a case, since for a Salpeter Initial Mass Function
(Salpeter 1955) 
we have that the distribution of stellar lifetimes behaves like
$\Psi(\tau_*) 
\sim \tau_*$ (since for main sequence stars,  $L \sim M^{3.45}$ e.g.\
Allen 
1973), it is immediately clear from Fig.~1, that it is {\it most
probable\/} 
that in the first place we encounter an intelligent civilization, we
will find 
that $\tau_l/\tau_* \sim 1$ (since the number of stars increases as we
move to 
the right in the diagram).  Therefore, if we can show that some
processes are 
likely to produce a monotonically increasing ($\tau_*$; $\tau_l/\tau_*$) 
relation, then the fact that $\tau_l \sim \tau_*$ in the earth-sun
system will 
find a natural explanation, and it will not have any implications for
the 
frequency of extraterrestrial civilizations.

I should note though that if the breadth of the `band' in Fig.~1 becomes
extremely large, this is essentially equivalent to no $\tau_l$--$\tau_*$ 
relation and Carter's argument is recovered. 

I will now give a simple example of how a $\tau_l$--$\tau_*$ relation
may arise. 
I should emphasize that this is not meant to be understood as a
realistic model, 
but merely to demonstrate that such a relation {\it could\/} exist.

Nucleic acid absorption of uv~radiation peaks between $2600$\AA\ and
$2700$\AA\ 
and that of proteins between $2700$\AA\ and $2900$\AA (e.g.\ Davidson
1960; 
Sagan 1961; Caspersson 1950).  Absorption in these bands is highly
lethal to 
all known forms of cell activity (e.g.\ Berkner 1952).  Of all the
potential 
constituents of a planet's atmosphere only O$_3$ absorbs efficiently in
the 
2000\AA--3000\AA\ range (e.g.\ Watanabe, Zelikoff \& Inn 1958).  It has
in fact 
been suggested that the appearance of land life has to await the
build-up of a 
sufficient layer of protective ozone (Berkner \& Marshall 1965; Hart
1978).  
Thus, it is important to understand the origin and evolution of oxygen
in 
planetary atmospheres.  While clearly only a limited knowledge of all
the 
processes involved exists (and even that only from the earth's
atmosphere), 
this will suffice for the purposes of the present example.  Two main
phases can 
be identified in the rise of oxygen in planetary atmospheres (Berkner \&
Marshall 
1965; Hart 1978; Levine, Hays \& Walker 1979; Canuto et~al.\ 1983). In
the first 
(which on earth lasted $\sim 2.4 \times 10^9$~yr), oxygen is released
from the 
photochemical dissociation of water vapor (this led on earth probably to
oxygen 
levels of $\sim 0.001$ of the present atmospheric level (P.A.L.)).  In
the 
second phase (which on earth lasted $\sim 1.6 \times 10^9$~yr), the
amounts of 
O$_2$ and O$_3$ reach levels $\sim 0.1$ P.A.L., sufficient to shadow the
land 
from lethal uv to allow the spread of life to dry land (the uv
extinction is 
normally expressed as $I_x = I_o~e^{-kx}$, where $I_o$ is the impinging 
intensity, $k$ is the absorption coefficient and $x$ is the path length
in the 
atmosphere). The important point to note (Berkner \& Marshall 1965; Hart
1978) 
is that the duration of the first phase is inversely proportional to the 
intensity of the radiation in the range  1000\AA--2000\AA\ (significant
peaks 
in H$_2$O absorption exist in the 1100\AA--1300\AA\ and
1600\AA--1800\AA\ ranges).  
Thus, for a given planetary size and orbit, the timescale for the
development 
of shielding (which we identify approximately with $\tau_l$) is
dependent on 
the stellar spectral type, and therefore on $\tau_*$.  For typical main
sequence 
star relations, $(L/L_{\odot}) = (M/M_{\odot})^{3.45}, (R/R_{\odot}) =
(M/M_{\odot})^\beta$ 
(with $\beta$ in the range 0.6--1, for spectral types F5--K5, see
below), and 
empirical fractions of the radiation emitted in the 1000\AA--2000\AA\
range 
(Stecker 1970; Carruthers 1971), a simple calculation (e.g.\ Livio \&
Kopelman 
1990) leads to an approximate relation of the form
\begin{equation}
\tau_l/\tau_* \simeq 0.4 (\tau_*/\tau^{\odot})^{1.7}~~~.
\end{equation}

\begin{figure}
\centerline{\epsfxsize=4in \epsfbox{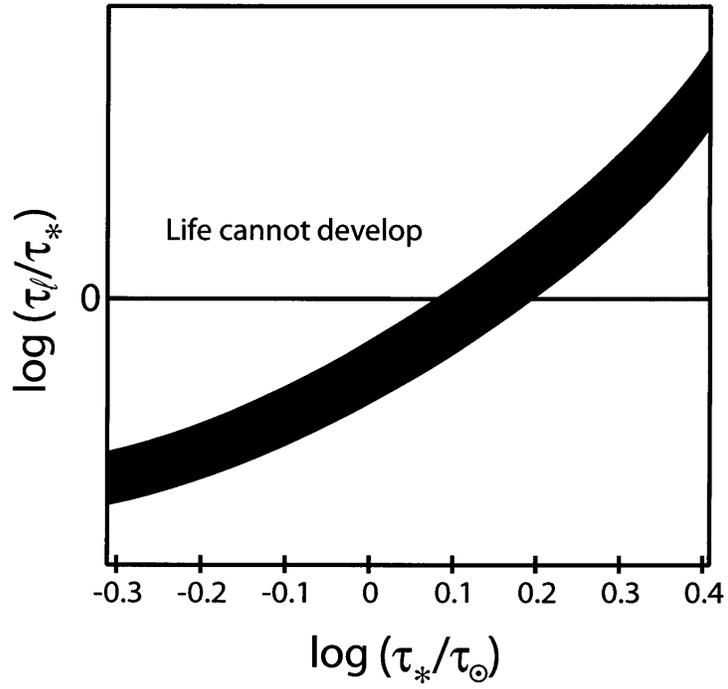}}
\caption{If the ratio of the timescale for biological evolution,
$\tau_l$, to 
the stellar lifetime, $\tau_*$, is a monotonically increasing function
of 
$\tau_*$, then the most likely relation is $\tau_l \sim \tau_*$ (see
text).}
\end{figure}
\begin{figure}
\centerline{\epsfxsize=4.6in \epsfbox{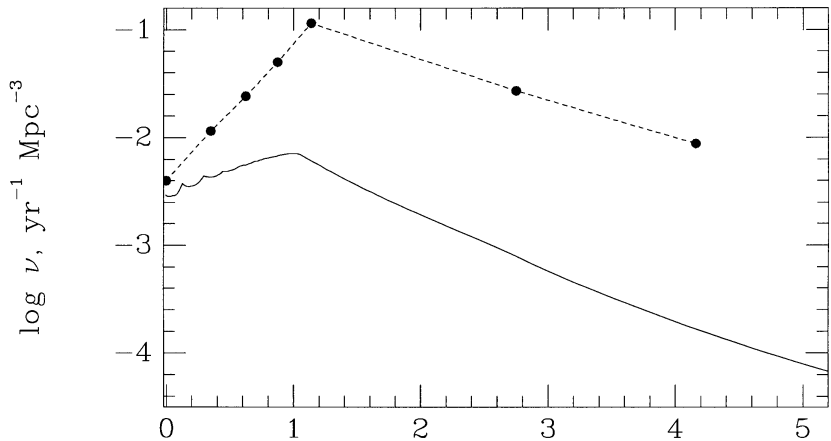}}
\caption{The dependence of the planetary nebulae formation rate on
redshift.  
Dashed curve---the star formation rate; solid line---planetary nebulae 
formation rate.}
\end{figure}
Clearly with the existence of a relation like~(1) (which is
monotonically 
increasing), the highest probability is to find $\tau_l \sim \tau_*$,
and hence 
the near equality of these two timescales in the earth-sun system cannot
be 
taken to imply that extraterrestrial civilizations are rare.

I should note again that the detailed evolution of the atmosphere is
surely more complicated than a simple dependence on the intensity of UV
photons. In particular, it may be that the different phases of the
evolution have different dependences on the properties of the central
star. The important point, however, is that, as the above example shows,
the {\it existence\/} of a $\tau_l$--$\tau_*$ relation is not
implausible, and that $\tau_l(\tau_*$) {\it could\/} increase faster
than linearly.

\section{When Did Intelligent Civilizations Emerge in the Universe?}

Given that extraterrestrial ``intelligent'' civilizations may not be 
exceedingly rare after all, one may ask what is a likely time in the
history 
of the universe for such civilizations to emerge.  I will restrict the 
discussion now to carbon-based civilizations.  Assuming a principle of 
`mediocrity', one would expect the emergence to coincide perhaps with
the peak 
in the carbon production rate. The main contributors of carbon 
to the interstellar medium are intermediate mass stars (Wood 1981;
Yungelson, 
Tutukov \& Livio 1993) through the Asymptotic Giant Branch (AGB) and
planetary 
nebulae (PNe) phases.  Recent progress has been made in the
understanding of 
the cosmic history of the star formation rate (SFR) (e.g.\ Madau et~al.\
1996; 
Lilly et~al.\ 1996; Madau, Pozzetti \& Dickinson 1997).  Assuming for 
simplicity that all the galaxies follow the same SFR history and stellar 
evolution processes, we can calculate the rate of formation of planetary 
nebulae (and hence the rate of carbon production) as a function of
redshift.  
For this purpose a population synthesis code which follows the evolution
of 
all the stars (assumed to be mainly in binaries), including all episodes
of 
mass exchange, common envelope phases etc., has been used (see
Yungelson, 
Tutukov \& Livio 1993; Yungelson et~al.\ 1996 for details of the code; I
am 
grateful to Lev Yungelson for carrying out the simulations).  In Fig.~2,
the 
assumed SFR as a function of redshift (taken as an approximation to the
results 
in Madau et~al.\ 1996) and the obtained PNe formation rate as a function
of 
redshift are shown.  As can be seen, the peak in the PN rate is somewhat 
delayed (to $z \simeq 1$) with respect to the peak in the SFR, and is
much more 
shallow at $z \aplt 1$, due to the build-up of a reservoir during the
previous 
epochs.  Realizing that continuously habitable zones (CHZs) exist only
around 
stars in the spectral range of about F5 to mid K (e.g.\ Kasting,
Whitmore \& 
Reynolds 1993), and that in general the biochemistry of life requires
rather 
precise conditions, carbon-based life may be expected to start (with the 
assumed SFR history) around $z \sim 1$, corresponding to an age of the
universe 
of $5.6 \times 10^9$~yr (for $\Omega_{0} = 0.2$, as seems to be
indicated by 
recent observations, e.g.\ Garnavich et~al.\ 1998; Parlmutter et~al.\
1998; 
Reiss, Press \& Kirshner 1996; Carlberg et~al.\ 1997, and a present age
of 
$t_0 = 13$ billion years).  Given the fact that (as I have shown in the 
previous section) the time required to develop intelligent civilizations
is 
$\tau_l \sim \tau_*$, it is expected that civilizations will emerge when
the 
age of the universe is $\apgt 10$~billion years, or maybe even somewhat
older, 
since the CHZs around K~stars are somewhat wider (in log distance) than
around 
G~stars.  A younger emergence age will be obtained if the star formation
rate 
does not decline at redshifts $1.2 \aplt z \aplt 5$, but rather stays
flat (as 
is perhaps suggested by the recent COBE Diffuse Background Experiment;
Hauser 
et~al.\ 1998; Calzetti \& Heckman 1998).

Finally, I should note that the arguments presented in this paper should 
definitely not be taken as attempting to imply that extraterrestrial
intelligent 
civilizations do exist. Rather, they show that the conclusion that they
don't, 
is at best premature (see also Rees 1997 for discussions of related
issues).

\begin{acknowledgments}
This work has been supported in part by NASA Grant NAG5-6857. I thank
the referee for his careful reading of the manuscript.
\end{acknowledgments}

\end{document}